\documentclass{IEEEtran}
\usepackage{cite}
\usepackage{amsmath,amssymb,amsfonts}
\usepackage{algorithmic}
\usepackage{graphicx}
\usepackage[table]{xcolor}
\usepackage{mathtools}
\usepackage{diagbox}
\usepackage{textcomp}
\usepackage{array}
\usepackage[multiple]{footmisc}
\usepackage{multirow} % <-- Required for \multirow table
\usepackage{textcomp}
\usepackage{xcolor}

\newtheorem{lemma}{Lemma}
\newtheorem{theorem}{Theorem}
\newtheorem{corollary}{Corollary}
\newtheorem{remark}{Remark}
\newtheorem{definition}{Definition}
\newtheorem{example}{Example}
\newtheorem{experiment}{Experiment}
\newtheorem{proposition}{Proposition}
\newtheorem{assumption}{Assumption}
\newtheorem{problem}{Problem}
\newtheorem{condition}{Condition}

\definecolor{mygray}{gray}{0.9}
\usepackage{tikz}
\usepackage{adjustbox}
\usetikzlibrary{calc}

\newcolumntype{M}[1]{>{\centering\arraybackslash}m{#1}}

\newtheorem{assumptionalt}{Condition}[assumption]
\newenvironment{assumptionp}[1]{
  \renewcommand\theassumptionalt{#1}
  \assumptionalt
}{\endassumptionalt}

\usepackage{dblfloatfix}
\def\BibTeX{{\rm B\kern-.05em{\sc i\kern-.025em b}\kern-.08em
    T\kern-.1667em\lower.7ex\hbox{E}\kern-.125emX}}

\usepackage{hyperref, comment}

\begin{document}
\title{Multi-Partite Output Regulation of \\ Multi-Agent Systems}

\author{K\"{u}r\c{s}ad Metehan G\"{u}l and Selahattin Burak Sars\i lmaz, \textit{Member, IEEE}
\thanks{K\"{u}r\c{s}ad Metehan G\"{u}l and Selahattin Burak Sars\i lmaz are with the Department of Electrical and Computer Engineering, Utah State University, Logan, UT 84322, USA (emails: 
        {\tt kursad.gul@usu.edu, burak.sarsilmaz@usu.edu}).}%
}

\maketitle
\begin{abstract}
This article proposes a simple, graph-independent perspective on partitioning the node set of a graph and provides multi-agent systems (MASs) with objectives beyond cooperation and bipartition. Specifically, we first introduce the notion of $k$-partition transformation to achieve any desired partition of the nodes. Then, we use this notion to formulate the multi-partite output regulation problem (MORP) of heterogeneous linear MASs, which comprises the existing cooperative output regulation problem (CORP) and bipartite output regulation problem (BORP) as subcases. The goal of the MORP is to design a distributed control law such that each follower that belongs to the same set in the partition asymptotically 
tracks a scalar multiple of the reference while ensuring the internal stability of the closed-loop system. It is shown that the necessary and sufficient conditions for the solvability of the MORP with a feedforward-based distributed control law follow from the CORP and lead to the first design strategy for the control parameters. However, it has a drawback in terms of scalability due to a partition-dependent condition. We prove that this condition is implied by its partition-independent version under a mild structural condition. This implication yields the second design strategy that is much more scalable than the first one. Finally, an experiment is conducted to demonstrate the MORP's flexibility, and two numerical examples are  provided to illustrate its generality and compare both design strategies regarding scalability.

%in accommodating shifting mission objectives. Two numerical examples are also provided to showcase its generality and compare both design strategies regarding scalability.
%\textcolor{blue}{an experimental demonstration and two} numerical examples are provided to illustrate the generality of the MORP and compare both design strategies regarding scalability. 
\end{abstract}


\section{Introduction}
\subsection{Motivation and Literature Review}
Considering the
studies on 
distributed control of multi-agent systems (MASs), two main frameworks become distinguishable: cooperative \cite{cai2022cooperative} and bipartite
\cite{6329411},\cite{Sekercioglu2025}. 
%\cite{6329411,Sekercioglu2025}.
%In both frameworks, consensus, synchronization,  leader-following consensus, formation, and output regulation stand out as the most recognized problems. 
The recognized problems of both frameworks include consensus
%, formation tracking,  
and output regulation. Regardless of the problem, MASs have a common objective in the cooperative framework, whereas they potentially have two opposed objectives in the bipartite framework.  

This article is motivated by the need for a flexible framework that extends beyond cooperation and bipartition within MASs to accommodate multiple, shifting mission objectives in adverse operating environments. 
For example, consider networked uninhabited aerial vehicles (UAVs) tasked with suppressing opponent air defense \cite{hoehn2022unmanned}. 
%In this scenario, 
During this operation, having the flexibility to achieve arbitrarily changing formations 
%throughout the operation 
is paramount to avoid radars while maximizing damage.
%\textcolor{blue}{For example, consider low-cost networked unmanned aerial vehicles (UAVs) tasked with a suppression and destruction of enemy air defense operation \cite{hoehn2022unmanned}. In this scenario, having the flexibility to achieve arbitrarily changing formations throughout the operation is of utmost importance to avoid radars while maximizing the damage.} 
Given that the output regulation problems for MASs 
allow  high-order nonidentical agent dynamics and 
%pave the way for a capability of tracking and rejecting a large class of signals, and
contain typical cooperative control problems, such as leader-following consensus,
%and formation tracking
we propose the multi-partite output regulation problem (MORP) to enhance tactical flexibility of MASs. The MORP includes the cooperative output regulation problem (CORP) and the bipartite output regulation problem (BORP) as special cases.

\begin{comment}
\textcolor{blue}{This article is mainly motivated by the need for a flexible framework that extends beyond cooperation and bipartition within MASs to accommodate multiple, shifting mission objectives, particularly in adverse operating environments. The aimed flexibility in objective definition is expected to make an impact on various fields, ranging from defense to commercial applications. Examples include suppression and destruction of enemy air defense operations \cite{hoehn2022unmanned}, wildfire management, and search and rescue.} 
\end{comment}  

Analogous to the output regulation problem, the CORP has been mainly treated using feedforward \cite{su2012cooperative, su2012cooperative_switch, 
cai2017adaptive, lu2017cooperativeMaobin, 10891994} and internal model \cite{kawamura2020distributed, sarsilmaz2021distributed, koru2020cooperative, koru2022regional, 10540292} approaches. In the former, the feedforward gain of each agent is based on a solution pair of the regulator equations, which are linear matrix equations (LMEs) determined by the exosystem and agent dynamics, making it not robust to parameter uncertainties. While the latter is known to be robust against small parameter variations,
it cannot be applied when the transmission zeros condition does not hold. 
\begin{comment}
    The studies with the feedforward approach can be divided into two: (i) \cite{su2012cooperative,su2012cooperative_switch,10891994,cai2017adaptive} use virtual information exchange (i.e., controller states of neighboring agents); (ii) \cite{lu2017cooperativeMaobin} uses only physical information exchange (i.e., relative outputs of neighboring agents). Similarly, one can classify the
articles solving
%articles addressing
the CORP via the internal model approach into two: (i) \cite{li2016distributed,7244394,kawamura2020distributed}
use virtual information exchange; (ii) \cite{sarsilmaz2021distributed, koru2020cooperative, koru2022regional,10540292} use only physical information exchange.
\end{comment}
Though fewer studies exist on the BORP, it is also tackled by the feedforward approach \cite{JiaoLewis, 8287187, 9089253} and the internal model approach \cite{huangbipartite}. 
In the CORP (BORP), the feedforward approach, with 
controller state exchange of neighboring agents, 
%virtual information exchange,
%where controller states of neighboring agents are exchanged,
%in which the exchanged information is the controller states of neighboring agents, 
uses a
distributed observer to provide the estimated state (estimated state or its additive inverse) of the exosystem to every agent.

Apart from the bipartite framework, there have been efforts to increase the number of objectives in MASs. 
The notable ones are cluster consensus 
%\textcolor{red}{on cooperative-competitive networks}
\cite{7944604,DEPASQUALE2022110002,eventtrig2}, scaled consensus \cite{ROY2015259}, and kernel manipulation of the Laplacian matrix 
%\cite{DeVries_Sims_Kutzer_2018, Tran17022025}. 
\cite{DeVries_Sims_Kutzer_2018},\cite{Tran17022025}.
%\textcolor{blue}{Event-triggered cluster consensus is also studied in \cite{eventtrig2}, \cite{eventtrig1}.} 
Yet, those studies are limited to consensus problems over first or second-order agent dynamics.
The cluster problem is also extended to high-order heterogeneous MASs  \cite{Chen01122020, optimal_cluster, 10737654}. The proposed MORP differs from the cluster consensus in two main aspects: First, the partitioning in the MORP is independent of the underlying graph, allowing each agent to determine its set in the partition, and hence, its objective, whereas the clustering in cluster consensus is graph-dependent, preventing each agent from self-determining its cluster. Second, while 
the MORP requires only one leader to generate multiple objectives, the cluster consensus requires at least two leaders to yield more than one objective. 

\subsection{Contribution}\label{sub:cont}
%This article formulates a general distributed control problem for heterogeneous MASs. To solve this problem, it provides two design strategies for a feedforward-based distributed control law involving a distributed observer with 
%comparable advantages and disadvantages. 

This article formulates a general distributed control problem for heterogeneous MASs. 
%This article formulates the MORP.
To solve this problem, it provides two design strategies with comparable advantages and disadvantages for a feedforward-based distributed control law. %involving a distributed observer.

%relying on virtual information exchange with comparable advantages and disadvantages. 

\begin{comment}
While the partition in the bipartite framework is graph-dependent,  we first introduce the notion of $k$-partition transformation to render the partition of the node set independent of the given graph. \textcolor{blue}{In fact, the proposed $k$-partition transformation allows arbitrary partitioning of the node set}. Hence, the node set can now admit up to $N$-partition, where $N$ is the cardinality of the node set. % In fact,  there are infinitely many $k$-partition transformations so that the node set can be partitioned the Bell number of ways. 
\end{comment}

The partition of the node set of a graph in the bipartite framework is graph-dependent (e.g., see Remark 2.1 and Equation (8) in \cite{huangbipartite}). In particular, each agent needs to know the graph's adjacency matrix to determine the set in which it lies in the bipartition, and hence, track the reference or its additive inverse.
To render the partition of the node set independent of the graph, we  introduce the notion of $k$-partition transformation. 
This not only allows the node set to 
admit up to $N$-partition, where $N$ is the cardinality of the node set, but also provides each agent with the flexibility to self-determine the set in which it lies in the $k$-partition, and hence,  which scalar multiple of the reference to track.

%This not only increases the number of objectives in the MAS up to the number of agents but also provides each agent with the flexibility to self-determine its objective. 

%Hence, the node set can now admit up to $N$-partition, where $N$ is the cardinality of the node set.

%Although the BORP increases the number of objectives to two, the bipartition of the agents depends on the graph topology (i.e., signs of the adjacency matrix).
   %This results in the following drawback.
%   In particular, each agent needs to know the graph's adjacency matrix to determine the set in which it lies in the bipartition, and hence, track the reference or its additive inverse. For example, see Remark 2.1 and equation (8) in  \textcolor{red}{reference [18]} of the revised manuscript. The definition of $\phi_i$ in Remark 2.1 depends on the graph's adjacency matrix through the graph-dependent bipartition of the node set. Correspondingly, the tracking error of each agent in (8) is graph-dependent. Hence, each agent cannot determine its objective among  two options (i.e., whether to track the reference or its additive inverse) in the BORP. 

By leveraging this notion, we formulate the MORP, which includes the CORP and BORP as special cases.
%(see Remark \ref{rmk:CompriseCORP} and Example \ref{exp:numexborp}).
To this end, we consider  a heterogeneous MAS in the form 
\begin{align}\label{eq:MAS}\nonumber
    \Dot{x}_i &= A_ix_i + B_iu_i + E_iv \\
    y_i &= C_ix_i+D_iu_i +G_iv, \quad i = 1,\dots,N
\end{align}
\noindent where $x_i \in \mathbb{R}^{n_i}$ is the state, $u_i \in \mathbb{R}^{m_i}$ is the control input, and $y_i \in \mathbb{R}^{p_i}$ is the output of subsystem $i$. Also,
$v \in \mathbb{R}^{n_0}$ is the exogenous signal generated by the following 
exosystem 
\begin{equation}\label{eq:exosystem}
    \Dot{v} = A_0v.
\end{equation}
This autonomous linear system yields the external disturbances $E_iv$ and $G_iv$ that the
state and output equations of subsystem $i$ are subject to, respectively,
and the reference denoted by $-F_iv$
for subsystem $i$.
%This autonomous linear system yields \textcolor{blue}{ the disturbances, which are $E_iv$ and $G_iv$, and the reference denoted by $-F_iv$
%for subsystem $i$.} 
%This autonomous linear system yields the reference (i.e., $-F_iv$) and disturbance (i.e., $E_iv$) signals for the MAS \eqref{eq:MAS}.  
The goal of the MORP is to design a distributed control law 
such that each subsystem in the same set within the partition asymptotically tracks 
a common
scalar multiple of the reference (i.e., a $k$-partition transformation term multiple of the reference) and rejects the disturbance while ensuring the internal stability of the closed-loop system.

Though the MORP can be cast as the CORP; hence, its necessary and sufficient solvability conditions follow from the CORP, the regulator equations become partition-dependent. %(see Condition \ref{ass:regeqsolution}).  
Consequently,
the immediate design strategy for control parameters, referred to as the first design strategy, requires each subsystem to recompute a solution pair for the regulator equations each time the $k$-partition transformation term is updated.  This is a drawback that does not exist in the CORP. 

An intriguing question arises from the drawback above: \emph{Is it possible to have a design strategy that eliminates the recomputation of a solution for the regulator equations whenever the $k$-partition transformation term changes?} 
Theorem \ref{thm:cnd3*tocnd3} paves the way for an affirmative answer under a mild structural condition. Specifically, it shows that the  solvability of the partition-independent regulator equations \eqref{eq:regulatoreqgeneral} and the introduced LME \eqref{eq:newcondition}, which solely depends on the subsystem data, ensures the existence of a solution pair to the partition-dependent regulator equations \eqref{eq:regulatoreqtriv}. 
More importantly, Theorem \ref{thm:cnd3*tocnd3} provides an analytical formula \eqref{eq:sol_pair_tilde} for such a pair in terms of the $k$-partition transformation term and the solutions of \eqref{eq:regulatoreqgeneral} and \eqref{eq:newcondition}. This leads to the second design strategy that does not require recomputing a solution to any equations when the $k$-partition transformation term changes.  
Accordingly, it is significantly more scalable than the first one. The discussion on the proposed design strategies' scalability and conservatism is summarized in Table \ref{tab:DesignStrategies} by referring to the corresponding conditions and results of the article. An experiment is conducted to demonstrate the MORP's flexibility in accommodating shifting mission objectives. 
Two numerical examples are also provided to showcase its generality and compare both design strategies regarding scalability.

\subsection{Notation}
The real part of a complex number $\lambda$ is denoted by $\mathrm{Re}(\lambda)$. The closed right (left) half complex plane is denoted by $\mathrm{CRHP}$ ($\mathrm{CLHP}$).
The open right half complex plane is denoted by $\mathrm{ORHP}$. 
We write $I_n$ for the $n \times n$ identity matrix,
$0_{n \times m}$ or $0$ for the $n \times m$ zero matrix, $\mathrm{diag} (w_1, \ldots, w_n)$ for  
the diagonal matrix with scalar entries $w_1, \ldots, w_n$ on its diagonal, and  $\otimes$ for the Kronecker product. 
%We denote the $n \times m$ zero matrix with $0_{n \times m}$ or $0$. 
%\textcolor{blue}{We denote $n \times m$ zero matrix with $0_{n \times m}$ unless clear from context; we use 0, otherwise.}
The spectrum of a square matrix 
$X \in \mathbb{R}^{n \times n}$ 
is denoted by $\mathrm{spec}(X)$. 
The matrix obtained by replacing each entry of $X$ with its absolute value is denoted by $|X|$.
The image of a matrix $Z \in \mathbb{R}^{n \times m}$
is denoted by $\mathrm{im}\hspace{0.03 cm}Z$.
%The vector formed by stacking columns of $Z$ is denoted by $\mathrm{vec}\hspace{0.03 cm}Z$. 
%The Kronecker product is 
%denoted by  $\otimes$. 

A (weighted) signed directed graph ${\mathcal{G}}$
is a triple ${\mathcal{G}} = (\mathcal{N}, \mathcal{E}, \mathcal{A})$, where 
$\mathcal{N} = \left \lbrace 1, \ldots, N \right \rbrace$ is the node set,
$\mathcal{E}  \subseteq \mathcal{N} \times \mathcal{N}$ is the edge set, and $\mathcal{A} = [a_{ij}] \in \mathbb{R}^{N\times N}$ is the adjacency matrix whose entries are determined by the rule: for any $j, i \in \mathcal{N}$, 
$a_{ij} \neq 0 $  if, and only if, $(j, i) \in \mathcal{E}$.
Graphs with self-loops are not considered in this article; that is, $a_{ii} = 0$ for $i=1,\ldots, N$. 
A graph is completely specified by its adjacency matrix $\mathcal{A}$, hence, the graph corresponding to $\mathcal{A}$ is denoted as $\mathcal{G}(\mathcal{A})$. 
A signed directed graph ${\mathcal{G}(\mathcal{A})}$ is called an unsigned directed graph if $a_{ij} \geq 0$ for $i,j = 1,\ldots,N$.
The in-degree and Laplacian matrices of a signed directed graph $\mathcal{G}(\mathcal{A})$, denoted by $\mathcal{D}$ and $\mathcal{L}$, are defined as $\mathcal{D} = \mathrm{diag}(d_{1},\ldots,d_{N})$ with $d_{i} = \sum_{j=1}^{N} |a_{ij}|$ for $i = 1,\ldots,N$ and $\mathcal{L} = \mathcal{D}-\mathcal{A}$. 

\section{Problem Formulation}\label{sec:Prob_Form}
This section introduces the notion of  $k$-partition transformation to achieve any desired partition for the node set of a given graph. It then uses this notion to formulate the MORP
of heterogeneous linear MASs.

\subsection{Arbitrary Partition of Nodes}\label{sec:arbitrarypartition}
We first recall the  $k$-partition of sets from combinatorics.

\begin{definition}[Sections 1.10, 5.1, and 5.4 in \cite{combinatorix}]\label{def:kpartition} 
    Let $T$ be a nonempty finite set of $N$ elements. A collection $\mathcal{T}$ of $1 \leq k \leq N$ nonempty subsets of $T$ is called a \emph{$k$-partition} of $T$ if 
    all the sets in 
    $\mathcal{T}$ are mutually disjoint and if their union equals $T$.
    The number of $k$-partitions of $T$ is called the \emph{Stirling number of the second kind}. The number of all partitions of $T$ is called the \emph{Bell number}. 
\end{definition}

Then, we introduce the $k$-partition transformation concept.  This   generalizes the gauge transformation (i.e., the signature matrix) used in  the bipartite framework \cite{6329411,Zhang2017,huangbipartite}.

\begin{definition}\label{def:transformation} 
A matrix $S = \mathrm{diag}(s_{1},\ldots,s_{N}) \in \mathbb{R}^{N\times N}$ with exactly $1 \leq k \leq N$ distinct entries on the diagonal is called a \emph{$k$-partition transformation}, where each $s_i$ is called a \emph{$k$-partition transformation term}. A 1-partition or 2-partition transformation $S$ is  a \emph{gauge transformation} if $s_i \in \{-1,1\}$ for $i=1, \ldots, N$. 
\end{definition}

Given a signed directed graph, the bipartite framework partitions the nodes according to the  signs of the adjacency matrix entries (i.e., the edge weights). This yields at most $2$-partition of nodes. In contrast, we partition the nodes according to a $k$-partition transformation.
Since a $k$-partition transformation can be chosen for any $k \in \{1, \ldots, N\} $,
the nodes can  admit not only 1-partition or 2-partition but also  more than 2-partition.
%\textcolor{blue}{\footnote{\textcolor{blue}{In the cluster problem with multiple leaders, though the clustering of the agents is graph-dependent due to the inter-cluster assumption as in \cite{optimal_cluster}, the $k$-partition of the agents within each cluster remains independent of the graph.}}}. 
%\textcolor{blue}{Owing to the fact $k$-partition transformation terms can be any real number, there are infinitely many $k$-partition transformations 
%so that the node set can be partitioned the Bell number of ways.}
%\textcolor{blue}{In fact, for each $k$-partition of the node set, there are infinitely many $k$-partition transformations because $k$-partition transformation terms can be any real number. 
%Hence, the Bell number can be achieved by infinitely many $k$-partition transformations.
%}
%\textcolor{blue}{On top of that, there are infinitely many $k$-partition transformations so that the node set can be partitioned the Bell number of ways.} 
In fact, there are infinitely many $k$-partition transformations so that the nodes can be partitioned in the Bell number of ways because $k$-partition transformation terms can be any real number.
We now formally discuss the $k$-partition of nodes. 

Let $\mathcal{G}(\mathcal{A})$ be a signed directed graph, and let $S$ be a $k$-partition transformation. Then, there exist 
$k$ positive integers
$i_1, \ldots, i_k $ such that $s_{i_1}, \ldots, s_{i_k}$ are $k$ distinct $k$-partition transformation terms. For each $p \in \left\{1, \ldots, k \right\}$, 
let %$   \mathcal{N}_{p} = \left\{
    %j \in \mathcal{N} ~|~  s_{j} = s_{i_p}
    %\right\}$.
%$\mathcal{N}_p$ denote the following set
\begin{eqnarray}
   \mathcal{N}_{p} = \left\{
    j \in \mathcal{N} ~|~  s_{j} = s_{i_p}
    \right\}.
\end{eqnarray}
Define the collection $\mathcal{C} = \left\{\mathcal{N}_1,\ldots,\mathcal{N}_k
\right\}$. 
Lemma \ref{lmm:$k$-partition} verifies that it is a $k$-partition of the node set $\mathcal{N}$.

\begin{lemma}\label{lmm:$k$-partition} 
The collection $\mathcal{C}$ has the following properties: 
\begin{enumerate}
        \item[(i)] If $\mathcal{N}_p \in \mathcal{C}$, then $\mathcal{N}_p \neq \emptyset$.
        \item[(ii)] 
        If $\mathcal{N}_p \in \mathcal{C}$ and $\mathcal{N}_r \in \mathcal{C}$ with $p\neq r$, then   $\mathcal{N}_p \cap \mathcal{N}_r = \emptyset$. 
        \item[(iii)] $\bigcup_{\mathcal{N}_p \in \mathcal{C}}\mathcal{N}_p = \mathcal{N}$.
    \end{enumerate} 
\end{lemma}
\begin{IEEEproof}
(i) Clearly, $i_p \in \mathcal{N}_p$. (ii) Let $\mathcal{N}_p \in \mathcal{C}$ and $\mathcal{N}_r \in \mathcal{C}$ with $p\neq r$, but assume for contradiction that $\mathcal{N}_p \cap \mathcal{N}_r \neq \emptyset$. Then, let $j \in  \mathcal{N}_p \cap \mathcal{N}_r$; hence, $j \in  \mathcal{N}_p$ and $ j \in \mathcal{N}_r$. By definition, $s_j = s_{i_p}$ and $s_j = s_{i_r}$. Thus, $s_{i_p} = s_{i_r} $, 
which contradicts $s_{i_p}$ and $s_{i_r}$ being distinct. (iii) Since every $\mathcal{N}_p \in \mathcal{C}$ is a subset of $\mathcal{N}$, the inclusion 
$\bigcup_{\mathcal{N}_p \in \mathcal{C}}\mathcal{N}_p \subseteq \mathcal{N}$ holds. 
Let  $j \in \mathcal{N}$.
Then, there exists a $p \in \left\{1, \ldots, k \right\}$ such that $s_j = s_{i_p}$ due to the fact that $S$ has exactly $k$ distinct diagonal entries. Hence, $j \in \mathcal{N}_p$ for some $\mathcal{N}_p \in \mathcal{C}$. This proves that  the inclusion 
$ \mathcal{N} \subseteq \bigcup_{\mathcal{N}_p \in \mathcal{C}}\mathcal{N}_p$ holds. 
\end{IEEEproof}
\begin{remark}\label{rmk:grpprobability}
Owing to Definition \ref{def:transformation} and Lemma \ref{lmm:$k$-partition}, the $k$-partition transformations can achieve arbitrary partition of nodes. For example, let $\mathcal{G}(\mathcal{A})$ be a %signed directed
graph with $5$ nodes. 
For $k=1,2,3,4,5$, there are infinitely many $k$-partition transformations that can obtain $1,15,25,10,1$ number of $k$-partitions of the nodes, respectively. These numbers correspond to the Stirling numbers of the second kind for respective values of $k$
(e.g., see Section 5.1 in \cite{combinatorix}). 
Their sum is $52$, 
which is the associated Bell number.
\end{remark}
\begin{comment}
\begin{remark}\label{rmk:grpprobability}
Owing to Definition \ref{def:transformation} and Lemma \ref{lmm:$k$-partition}, $k$-partition transformations can achieve arbitrary partition of nodes. For example, let $\mathcal{G}(\mathcal{A})$ be a signed directed graph with $5$ nodes. 
For $k=1,2,3,4,5$, there exist infinitely many $k$-partition transformations  that can be used to define $1,15,25,10,1$ number of $k$-partitions of the node set, respectively.   These numbers correspond to the Stirling numbers of the second kind for respective values of $k$
(e.g., see Section 5.1 in \cite{combinatorix}). 
Their sum is $52$, which is the associated Bell number.
\end{remark}
\end{comment}

\subsection{MORP}\label{subsec: MORP_graph}

As in the context of the CORP and BORP,  the subsystems of  \eqref{eq:MAS}, considered followers, and the exosystem \eqref{eq:exosystem}, considered the leader,  constitute a MAS of $N+1$ agents. 
To model the information exchange between $N$ followers, we use a time-invariant signed directed graph ${\mathcal{G}}(\mathcal{A})$ without self-loops as  $\mathcal{N} = \left \lbrace 1, \ldots, N \right \rbrace$, where node $i \in \mathcal{N}$ corresponds to follower $i$, and for each $j,i \in \mathcal{N}$, we put $(j,i) \in \mathcal{E}$ if, and only if, follower $i$ has access to the information of follower $j$.  The leader is included in the information exchange model by augmenting the graph ${\mathcal{G}}(\mathcal{A})$. Specifically, let  $\mathcal{G}(\Bar{\mathcal{A}})$ be an augmented signed directed graph with  $\Bar{\mathcal{N}} = \mathcal{N} \cup \{0\}$, $\Bar{\mathcal{E}} = \mathcal{E} \cup \mathcal{E}'$, where $\mathcal{E}' \subseteq \{(0,i) \  |  \ i \in \mathcal{N}\}$, and $\Bar{\mathcal{A}} \in \mathbb{R}^{(N+1)\times (N+1)}$. Here, node $0$ corresponds to the leader and for any $i \in \mathcal{N}$, we put  $(0,i) \in \mathcal{E}'$ if, and only if, follower $i$ has access to the information of the leader. For any $i\in \mathcal{N}$, the pinning gain $f_{i} > 0$ if $(0,i) \in \mathcal{E}'$ and $f_{i} = 0$ otherwise.
The pinning gain matrix
is defined by 
$\mathcal{F} = \mathrm{diag}(f_{1},\ldots,f_{N})$. 

A control law that relies on the information exchange modeled by an augmented signed directed graph $\mathcal{G}(\Bar{\mathcal{A}})$ is called a \emph{distributed control law}. The \emph{closed-loop system} consists of \eqref{eq:MAS} and the distributed controller. We now formulate the MORP. 

\begin{problem}[MORP]\label{prb:mainprob}
    Given the heterogeneous MAS composed of \eqref{eq:MAS} and \eqref{eq:exosystem}, an augmented signed directed graph $\mathcal{G}(\Bar{\mathcal{A}})$, and a  $k$-partition transformation $S$, 
    find a distributed control law  such that
\begin{enumerate}
    \item[(i)] The closed-loop system matrix is Hurwitz;
    \item[(ii)] For any initial state of the exosystem and closed-loop system, the tracking error of each $i \in \mathcal{N}$ defined by   
    \begin{eqnarray}\label{eq:tracking_error}
        e_i = y_i + s_{i}F_iv   
    \end{eqnarray}
    satisfies
 $\lim_{t\to\infty} e_i(t) = 0$. 
 \end{enumerate}
\end{problem}
\begin{remark}\label{rmk:CompriseCORP}
The MORP includes  the  CORP and BORP objectives:
If $S = I_N$,  the MORP reduces to the CORP (e.g., see Definition 1 in \cite{su2012cooperative}).  If $S$ is a gauge transformation,  the MORP reduces to the BORP (e.g., see Problem 2.1 in \cite{huangbipartite}).
\end{remark}

\section{Solvability of the MORP}\label{sec:MORPClassicalApproach}
This section first observes
that the MORP can be realized as the CORP. 
We then consider 
one of the feedforward-based distributed control laws solving the CORP.
Based on this control law and the observation, necessary and sufficient conditions for the solvability of the MORP follow from the CORP, 
resulting in the first design strategy for control parameters.
Though this strategy is straightforward, it has a drawback due to the $k$-partition transformation dependence of the regulator equations. 
The section concludes
with a discussion of this drawback. 

\subsection{Distributed Control Law}

Using $\eqref{eq:tracking_error}$ and defining $\tilde{F}_i = G_i + s_iF_i$ for each $i \in \mathcal{N}$, we can rewrite \eqref{eq:MAS} as
\begin{align}\label{eq:MAS_rewrite}\nonumber
    \Dot{x}_i &= A_ix_i + B_iu_i + E_iv \\
    e_i &= C_ix_i+D_iu_i +\tilde{F}_iv, \quad i = 1, \ldots, N.
\end{align}
Then, the following result is immediate. 

\begin{lemma}\label{lmm:observation}
 Any 
 %linear time-invariant 
 distributed control law solving the CORP for the MAS composed of \eqref{eq:MAS_rewrite} and \eqref{eq:exosystem}  over an augmented unsigned directed graph solves the MORP over the same graph. 
\end{lemma}

In the light of Lemma \ref{lmm:observation},
we consider the following feedforward-based distributed control law\footnote{With the distributed measurement output feedback control law in Equation (8.14) of \cite{cai2022cooperative}, one can  arrive at a result similar to Theorem \ref{thm:trivialtheorem} (ii) under the additional detectability assumptions. The resulting design strategy will have the drawback in Remark \ref{rmk:discussion_drawback}. In this case, Theorem \ref{thm:cnd3*tocnd3} is still a remedy.}\textsuperscript{,}\footnote{Each follower is assumed to access $A_0$.  To relax this assumption, the adaptive distributed observer, adaptive solution of regulator equations, and corresponding time-varying feedforward gain in \cite{cai2017adaptive} can be employed to arrive at a result similar to Theorem \ref{thm:trivialtheorem} (ii). Since the adaptive solution of regulator equations  will be partition-dependent, there will be a transient response due to the adaptation when $s_i$ is updated. 
%the transient response will be affected whenever $s_i$ is updated. 
In this case, Theorem \ref{thm:cnd3*tocnd3} helps eliminate the partition-dependence, and hence, the associated transient response.} proposed in 
\cite{su2012cooperative}
\begin{eqnarray}\label{eq:classicalcontroller}\nonumber
    \dot \eta_i \hspace{-3mm}&=& \hspace{-3mm} A_0\eta_i + \mu \Big(\sum_{j=1}^{N}  |a_{ij}| (\eta_j-\eta_i) +f_i(v-\eta_i) \Big) \\
    u_i \hspace{-3mm}&=& \hspace{-3mm} K_{1i}x_i + K_{2i}\eta_i, \quad i = 1,\ldots,N
\end{eqnarray}
where $\eta_i \in \mathbb{R}^{n_0}$ is 
the estimate 
% the exogenous signal
of $v$.
The state equation in \eqref{eq:classicalcontroller} is called a \emph{distributed observer}.
Moreover, $\mu \in \mathbb{R}$, 
$K_{1i} \in \mathbb{R}^{m_i \times n_i}$, and $K_{2i} \in \mathbb{R}^{m_i \times n_0}$ are control parameters to be designed. 
Compared to the distributed observer in 
\cite{su2012cooperative}, the one in \eqref{eq:classicalcontroller} uses the absolute value of the adjacency matrix entries to be readily applicable even if the given graph is signed. 

\subsection{MORP: Necessary and Sufficient Conditions}

The following conditions will be referred to
for the solvability of Problem \ref{prb:mainprob}.

\begin{condition}\label{ass:SantiHurwitz}
The inclusion $\mathrm{spec}(A_0) \subsetneq \mathrm{CRHP}$ holds.
\end{condition}
\begin{condition}\label{ass:ABStabilizable} 
For any $i \in \mathcal{N}$, the pair $(A_i,B_i)$ is stabilizable. 
\end{condition}
\begin{condition}\label{ass:regeqsolution} 
For any $i \in \mathcal{N}$, there exists a pair $(X_i,U_i)$ that satisfies the regulator equations 
\begin{eqnarray}\label{eq:regulatoreqtriv}\nonumber
    X_iA_0 &=& A_iX_i + B_iU_i + E_i\\ 
    0 &=& C_iX_i + D_iU_i + G_i + s_{i} F_i.
\end{eqnarray}
\end{condition}
\begin{condition}\label{ass:spngtree}
The augmented signed directed graph $\mathcal{G}(\Bar{\mathcal{A}})$ contains a directed spanning tree{\footnote{By the definition of augmented signed directed graphs given in Section \ref{subsec: MORP_graph}, the root of any directed spanning tree in $\mathcal{G}(\Bar{\mathcal{A}})$ is necessarily node $0$.
}}.

%\textcolor{blue}{\footnote{\textcolor{blue}{The root node of such a spanning tree is necessarily the leader node.}}}

\end{condition}

\begin{remark}\label{rmk:assumptions}
Conditions \ref{ass:SantiHurwitz}, \ref{ass:ABStabilizable}, and \ref{ass:spngtree} are standard while tackling both the CORP and BORP (e.g., see \cite{su2012cooperative, 9089253}).
The BORP literature imposes the structural balance condition
on either  $\mathcal{G}(\Bar{\mathcal{A}})$ (e.g.,  see Assumption 3.1 in \cite{JiaoLewis}) or $\mathcal{G}(\mathcal{A})$ (e.g., see Assumption 5 in  \cite{9089253}). Yet, this article does not require such a condition because the distributed observer in \eqref{eq:classicalcontroller} uses the absolute value of the adjacency matrix entries.   
\end{remark}

\begin{remark}
The $k$-partition transformation term $s_i$ is incorporated into \eqref{eq:regulatoreqtriv} due to Lemma \ref{lmm:observation}.
Except for this term, Condition \ref{ass:regeqsolution} 
is standard for the studies investigating the CORP with the feedforward approach
(e.g., see \cite{su2012cooperative, lu2017cooperativeMaobin}). 
The articles \cite{9089253, JiaoLewis} studying the BORP with the feedforward approach consider the disturbance-free (i.e., $E_i=0$ 
and $G_i = 0$)  and direct feedthrough-free (i.e., $D_i=0$) follower dynamics. 
However, introducing a gauge transformation term in the regulator equations for each follower can extend their distributed controllers to take the effect of disturbances and direct feedthrough into account. To  avoid such modification, the authors in \cite{8287187} assume that agents in different sets in partition are subject to disturbances that are additive inverses of each other. This can be impractical in real-world applications, for instance, consider networked UAVs operating in the same environment. Thus, this article does not make that assumption about disturbances.
\end{remark}

We now provide the necessary and sufficient conditions for the solvability of the MORP. 
 \begin{theorem}\label{thm:trivialtheorem}
The following statements are true: 
\begin{enumerate}
    \item[(i)] Suppose Condition \ref{ass:SantiHurwitz} holds. If Problem \ref{prb:mainprob} is solvable by the distributed control law \eqref{eq:classicalcontroller}, then Conditions \ref{ass:ABStabilizable}--\ref{ass:spngtree} hold. 
    \item[(ii)] Under Conditions \ref{ass:ABStabilizable}--\ref{ass:spngtree}, Problem \ref{prb:mainprob} is solvable by the distributed control law \eqref{eq:classicalcontroller}. 
\end{enumerate}
\end{theorem}
\begin{IEEEproof}
  The proof can be conducted by following the procedure in the proof of Theorem 1 in \cite{10644520}.
\end{IEEEproof}

\begin{remark}
Conditions \ref{ass:ABStabilizable}--\ref{ass:spngtree} are sufficient
 for the solvability of the MORP by the distributed control law \eqref{eq:classicalcontroller}. Under 
 Condition \ref{ass:SantiHurwitz}, they are also necessary.
 %Conditions \ref{ass:ABStabilizable} and \ref{ass:spngtree} are also necessary under Condition \ref{ass:SantiHurwitz}.  
\end{remark}

\subsection{Discussion: Synthesis of Control Parameters} 
This subsection presents the first design strategy for the control parameters and highlights an associated drawback. 
%To this end, 
Let $\mathcal{L}_{u}$ denote the Laplacian matrix of the unsigned directed graph $\mathcal{G(|\mathcal{A}|)}$ and $\mathcal{H} = \mathcal{L}_u + \mathcal{F}$. 
Let $\lambda_i(\mathcal{H})$ and 
$\lambda_j(A_0)$ denote the eigenvalues of $\mathcal{H}$ and 
$A_0$ for 
%$i \in \mathcal{N}$ and
$i = 1, \ldots, N$,
$j = 1, \ldots, n_0$. 

\begin{remark}\label{rmk:designtrivial} Let Conditions \ref{ass:ABStabilizable}--\ref{ass:spngtree} hold. 
 The constructive procedure in the proof of  Theorem  \ref{thm:trivialtheorem} (ii) yields the following design steps for the parameters $\mu$, $K_{1i}$, and $K_{2i}$:
 \begin{enumerate}
\item[(i)] Select $\mu$ according to the inequality\footnote{Ensuring the Hurwitzness of the matrix $A_{\mu} = (I_N \otimes A_0) - \mu(\mathcal{H} \otimes I_{n_0})$, the system matrix of the distributed observer in compact form, is crucial in the proof of Theorem \ref{thm:trivialtheorem} (ii). Specifically, Lemma 4 (ii) of \cite{10644520} gives a sufficient but not necessary condition for the  Hurwitzness of $A_{\mu}$ under Condition \ref{ass:spngtree}, and it ignores instances where a nonpositive $\mu$ can still make $A_{\mu}$ Hurwitz. However, Lemma \ref{lmm:muvalue} in Appendix characterizes the Hurwitzness of $A_{\mu}$ through the inequality \eqref{eq:muselection} under Condition \ref{ass:spngtree}. Thus, the lower bound on $\mu$ in \eqref{eq:muselection} has no conservatism compared to the one in Lemma 4 of \cite{10644520}. Another lower bound is also provided in Lemma 
3.2 of \cite{cai2022cooperative}. But, one can construct a counterexample to the first statement of that lemma by considering a Hurwitz $A_0$.}

\begin{equation}\label{eq:muselection}
            \mu > 
            \max_{\substack{ j \in \{1, \ldots, n_0 \hspace{-0.03 cm}\} \\ i \in  \{1, \ldots, N\hspace{-0.01 cm}\}}}
       \frac{\mathrm{Re}(\lambda_{j}(A_0))}{\mathrm{Re}(\lambda_{i}(\mathcal{H}))}. 
        \end{equation}  
\item[(ii)] For each $i \in \mathcal{N}$, design $K_{1i}$ such that $A_i+B_iK_{1i}$ is Hurwitz. 
     \item[(iii)] For each $i \in \mathcal{N}$, find a pair $(X_i,U_i)$ that satisfies the regulator equations \eqref{eq:regulatoreqtriv}. 
          \item[(iv)] For each $i \in \mathcal{N}$, let $K_{2i} = U_i-K_{1i}X_i$.      
 \end{enumerate}   
As $\mathrm{spec}(\mathcal{H}) \subsetneq \mathrm{ORHP}$ under Condition \ref{ass:spngtree} (e.g., see Lemma 1 in \cite{su2012cooperative}),  
any positive $\mu$ satisfies the inequality \eqref{eq:muselection} if $\mathrm{spec}(A_0) \subsetneq \mathrm{CLHP}$. 
Thus, the design of $\mu$ is independent of the eigenvalues of $\mathcal{H}$ (i.e., the spectral property of  $\mathcal{G}(\Bar{\mathcal{A}})$)
when the exosystem  generates a linear  combination of  
constant signals, sinusoidal signals, polynomial signals, for instance, ramp signals, and exponentially converging signals. These signals cover references and disturbances encountered in most MASs (e.g., see the applications in \cite{Sekercioglu2025,su2012cooperative_switch,10891994,Sarsılmazbookchapter}).
\end{remark}

\begin{remark}\label{rmk:discussion_drawback}
Though the design based on Remark \ref{rmk:designtrivial} is straightforward to apply, it has a drawback in that the pair  $(X_i,U_i)$ for each follower depends on  $s_i$. 
Therefore, whenever the $k$-partition transformation is updated, it necessitates each follower to recompute a solution pair for the regulator equations \eqref{eq:regulatoreqtriv} unless its $s_i$ remains unchanged. Hence, given a finite set of $k$-partition transformation terms with $M$ elements, each follower computes $M$ solution pairs to \eqref{eq:regulatoreqtriv}.
This drawback can render the design strategy impractical in some
applications. 
For example, consider low-cost networked  UAVs trying to avoid radars. In this scenario, the set of $k$-partition transformation terms for each follower may not be known before the operation. Thus, it may be desirable in terms of computational cost to find a solution pair for the regulator equations \eqref{eq:regulatoreqtriv} once and use it throughout the operation. 
\end{remark}

The following section, motivated by the discussion in Remark \ref{rmk:discussion_drawback}, seeks an answer to the question posed in Section \ref{sub:cont}. 

\section{MORP: Partition-Independent Solvability}

This section first provides the partition-independent sufficient conditions for the solvability of the MORP.  For each $i \in \mathcal{N}$, these conditions include the partition-independent regulator equations, as in the CORP with the feedforward approach, and an LME depending only on $B_i$, $E_i$, $D_i$, and $G_i$. 
The constructive nature of the result yields the second design strategy 
that 
eliminates 
%does not suffer from
 the drawback discussed in  Remark \ref{rmk:discussion_drawback}.
Lastly, we reveal that the introduced LME's solvability is equivalent to an easily testable mild structural condition.  
%Then, we reveal that the solvability of the LME is equivalent to a mild structural condition, which is easily testable. 
%The section concludes with a system of linear equations that can be used to recover a solution to the LME.

\subsection{MORP: Partition-Independent Sufficient Conditions}
We modify Condition \ref{ass:regeqsolution} by removing the $k$-partition transformation term in \eqref{eq:regulatoreqtriv}.
\begin{assumptionp}{\ref*{ass:regeqsolution}$^*$}\label{ass:cnd3*}
For any $i \in \mathcal{N}$, there exists a pair $(X_i,U_i)$ that satisfies the partition-independent regulator equations
\begin{eqnarray}\label{eq:regulatoreqgeneral}\nonumber
    X_iA_0 &=& A_iX_i + B_iU_i + E_i\\ 
    0 &=& C_iX_i + D_iU_i + G_i + F_i.
\end{eqnarray}
\end{assumptionp}

Theorem \ref{thm:cnd3*tocnd3}
not only shows that the solvability of the 
partition-independent regulator equations \eqref{eq:regulatoreqgeneral} and  the LME 
\eqref{eq:newcondition} ensures the solvability of the partition-dependent regulator equations  \eqref{eq:regulatoreqtriv} but also provides a solution pair. 

\begin{theorem}\label{thm:cnd3*tocnd3}
If Condition \ref{ass:cnd3*} holds and if, for any $i\in \mathcal{N}$, there exists a
solution to the following LME
\begin{eqnarray}\label{eq:newcondition}
    \begin{bmatrix}
        B_i \\
        D_i
    \end{bmatrix} Y_i = \begin{bmatrix}
        E_i \\ G_i
    \end{bmatrix}
\end{eqnarray}
then Condition \ref{ass:regeqsolution} holds. In particular, for any $i\in \mathcal{N}$,  the  pair $(\tilde X_i, \tilde U_i)$ given by
\begin{eqnarray}\label{eq:sol_pair_tilde}
    \tilde X_i &=& s_i X_i \nonumber  \\ 
    \tilde U_i &=& s_i(U_i + Y_i) - Y_i
\end{eqnarray}
satisfies the partition-dependent regulator equations \eqref{eq:regulatoreqtriv}.
 
\end{theorem}
\begin{IEEEproof}
Fix $i\in \mathcal{N}$. Let $(X_i, U_i)$ be a pair that satisfies the partition-independent regulator equations \eqref{eq:regulatoreqgeneral}. Also, let $Y_i$ be
a solution to the LME \eqref{eq:newcondition}. 
By the pair $(\tilde X_i, \tilde U_i)$  in \eqref{eq:sol_pair_tilde},  
\begin{align}\label{eq:new_pair_solving_1}
  A_i \tilde X_i + B_i \tilde U_i +E_i &=
    s_i (A_i X_i + B_iU_i + B_i Y_i) \hspace{-0.02 cm}- \hspace{-0.02 cm}B_i Y_i +E_i \nonumber \\
  &= s_i(A_i X_i + B_iU_i +E_i)  \nonumber \\ &= s_i X_i A_0 = \tilde X_i A_0
    \end{align}
where the second equation follows from 
$B_iY_i = E_i$, the third equation is due to the fact that $  A_iX_i + B_iU_i + E_i = X_iA_0  $, and the fourth equation is a consequence of $\tilde X_i = s_i X_i$.  By the pair $(\tilde X_i, \tilde U_i)$  in \eqref{eq:sol_pair_tilde},  we further have
%\begin{align}\label{eq:new_pair_solving_2}
%  C_i \tilde X_i \hspace{-0.02cm}+\hspace{-0.02cm} D_i \tilde U_i \hspace{-0.02cm}+ \hspace{-0.02cm}s_iF_i &=
%    s_i \left(C_i X_i \hspace{-0.02cm}+\hspace{-0.02cm} D_i(U_i \hspace{-0.02cm}+\hspace{-0.02cm} Y_i) +F_i\right) \hspace{-0.03cm}-\hspace{-0.03cm} D_i Y_i  \nonumber \\
%  &= s_i(C_i X_i + D_iU_i +F_i) = 0
%    \end{align}
\begin{align}\label{eq:new_pair_solving_2}
 \hspace{-0.cm} & \hspace{0.1 cm} C_i \tilde X_i \hspace{-0.02cm}+\hspace{-0.02cm} D_i \tilde U_i \hspace{-0.02cm}+\hspace{-0.02cm} G_i \hspace{-0.02cm}+ \hspace{-0.02cm}s_iF_i \nonumber 
\\
 \hspace{-0.cm} = & \hspace{0.1 cm}s_i \left(C_i X_i \hspace{-0.02cm}+\hspace{-0.02cm} D_i(U_i \hspace{-0.02cm}+\hspace{-0.02cm} Y_i) +F_i\right) \hspace{-0.03cm}-\hspace{-0.03cm} D_i Y_i + G_i \nonumber  \\
 \hspace{-0.cm} = & \hspace{0.1 cm}  s_i(C_i X_i + D_iU_i + G_i + F_i) = 0
    \end{align}
where the second equation follows from $D_iY_i = G_i$ and the third equation is a result of $C_iX_i + D_iU_i + G_i+F_i = 0$. We now conclude from \eqref{eq:new_pair_solving_1} and \eqref{eq:new_pair_solving_2} that the pair $(\tilde X_i, \tilde U_i)$  in \eqref{eq:sol_pair_tilde} satisfies the partition-dependent regulator equations \eqref{eq:regulatoreqtriv}. 
%Hence, the proof is over. 
%Hence the proof.
\end{IEEEproof}

\begin{remark}\label{rmk:converse}
The converse of the first statement in Theorem \ref{thm:cnd3*tocnd3} is not true. To show this, consider a MAS consisting of one leader and one follower with system parameters $A_0 = G_1 = 0$, $A_1 = B_1 = C_1 = D_1 = 1$, $E_1 = -2$, $ F_1 = -1$, and $k$-partition transformation term $s_1 = 2$. It can be
%easily
verified that the pair $(1,1)$ satisfies the partition-dependent regulator equations \eqref{eq:regulatoreqtriv}.  We assume for contradiction that there is a pair $(X_1,U_1)$ that satisfies the partition-independent regulator equations \eqref{eq:regulatoreqgeneral} and a solution $Y_1$ to the LME \eqref{eq:newcondition}.  Then, \eqref{eq:regulatoreqgeneral} yields $X_1 + U_1 = 2$ and $X_1 + U_1 = 1$. Hence, $1=2$, a contradiction. We have just proved that the converse of the first statement in Theorem  \ref{thm:cnd3*tocnd3} is not true. One can also show that the LME \eqref{eq:newcondition} does not have a solution for the considered example.  
\end{remark}

In conjunction with  Theorem \ref{thm:trivialtheorem} (ii), Theorem \ref{thm:cnd3*tocnd3} leads to the following partition-independent sufficient conditions for the solvability of the MORP. 

\begin{corollary}\label{crl:maintheorem}
Under Conditions \ref{ass:ABStabilizable}, \ref{ass:cnd3*}, and \ref{ass:spngtree}, 
 Problem \ref{prb:mainprob} is solvable by the distributed control law \eqref{eq:classicalcontroller} if, for any $i\in \mathcal{N}$, there is a solution $Y_i$ to the LME \eqref{eq:newcondition}.
\end{corollary}

The rest of this subsection 
recalls a well-known alternative sufficient condition for  Condition \ref{ass:regeqsolution},  highlights the importance of Theorem \ref{thm:cnd3*tocnd3} in design, and compare the sufficient conditions. 

\begin{remark}\label{rmk:transzero}
Alongside Theorem \ref{thm:cnd3*tocnd3},
we know from Theorem 1.9 in \cite{huang2004nonlinearbook} that another partition-independent sufficient condition for Condition \ref{ass:regeqsolution}  is that, for any $i\in \mathcal{N}$, the rank condition\footnote{It is known as the transmission zeros condition (e.g., see Remark 1.11 in \cite{huang2004nonlinearbook}) if
the pair $(A_i,B_i)$  is controllable and the pair $(A_i,C_i)$ is observable.} 
\begin{eqnarray}\label{eq:transmission_zeros}
         \mathrm{rank}\begin{bmatrix}
               A_i - \lambda_j(A_0) I_{n_i}& B_i \\
                C_i & D_i 
            \end{bmatrix} = n_i + p_i, \  j = 1, \ldots, n_0
        \end{eqnarray}
        holds.  
Similar to the conditions in Theorem \ref{thm:cnd3*tocnd3},
this sufficient condition is not necessary for Condition \ref{ass:regeqsolution} to hold. To see this, consider the example in Remark \ref{rmk:converse}.         
What significantly distinguishes Theorem \ref{thm:cnd3*tocnd3} from this sufficient condition is that  Theorem \ref{thm:cnd3*tocnd3}
provides a solution pair, given by \eqref{eq:sol_pair_tilde}, to the partition-dependent regulator equations \eqref{eq:regulatoreqtriv}. This pair is explicitly expressed in terms of the $k$-partition transformation term and the matrices satisfying the partition-independent regulator equations \eqref{eq:regulatoreqgeneral} and the LME \eqref{eq:newcondition}.  Thus, in the second design strategy to be given,
each follower can recompute its feedforward gain $K_{2i}$ without recomputing a solution to any  equations when the $k$-partition transformation term is updated.
\end{remark}

\begin{remark}
    Consider a MAS consisting of one leader and one follower.  Let $\Phi$ (respectively, $\Theta$) be the set of leader and follower parameters that satisfy the conditions in Theorem \ref{thm:cnd3*tocnd3} (respectively, \eqref{eq:transmission_zeros}). Then, we have
        \begin{eqnarray}
        \Phi \setminus \Theta \neq \emptyset, \quad \Theta \setminus \Phi \neq \emptyset, \quad \Phi \cap \Theta \neq \emptyset.
    \end{eqnarray}
 To see this,  we first consider  the following system parameters \begin{eqnarray}\nonumber
            A_0 = \begin{bmatrix}
                0 & 1 \\ -1 & 0
            \end{bmatrix}, A_1 = \begin{bmatrix}
                0 & 1 \\ 0 & 0
            \end{bmatrix}, B_1 = \begin{bmatrix}
                0 \\ 1
            \end{bmatrix}, E_1 = \begin{bmatrix}
                0 & 0 \\ 0 & 0.5
            \end{bmatrix}
    \end{eqnarray}
$C_1 = I_2$, $D_1 = 0$, $G_1 = 0 $, and $F_1 = -I_2$.  
Observe that the conditions in Theorem \ref{thm:cnd3*tocnd3} hold with $X_1 = I_2$, $U_1 = - [1 \ 0.5]$ and $Y_1 = [0 \ 0.5]$, 
but the condition in \eqref{eq:transmission_zeros} does not hold. Hence, the system parameters belong to $\Phi \setminus \Theta$. Second, we consider 
the example in Remark  \ref{rmk:converse} with $A_0 = 1$. 
The condition in \eqref{eq:transmission_zeros} holds, but there is no 
$Y_1$ solving the LME \eqref{eq:newcondition}.  
Thus, the system parameters belong to $\Theta \setminus \Phi$. Third, we consider the example in  Remark  \ref{rmk:converse} with $A_0 = 1$ and $E_1 = 0$. The conditions in Theorem \ref{thm:cnd3*tocnd3} hold with $X_1 = 1$, $U_1 = 0$, and $Y_1 = 0$. The condition in \eqref{eq:transmission_zeros} also holds. Hence,  the system parameters belong to $\Phi \cap \Theta$. We have shown that 
the intersection of $\Phi$ and $\Theta$ is nonempty, and none of each is a subset of the other.
Therefore, exploring a  solution pair structure to the partition-dependent regulator equations \eqref{eq:regulatoreqtriv} analogous to Theorem \ref{thm:cnd3*tocnd3} under the sufficient condition in \eqref{eq:transmission_zeros} may expand the set of leader and follower parameters for which the second design strategy, given in the following subsection, is applicable.

%Thus, it is well worth exploring a solution pair structure to the partition-dependent regulator equations \eqref{eq:regulatoreqtriv} analogous to Theorem \ref{thm:cnd3*tocnd3} under the sufficient condition in \eqref{eq:transmission_zeros}. If successful, it expands the set of leader and follower parameters for which the second design strategy, given in the following subsection, is applicable.  
\end{remark}

\subsection{Discussion: Synthesis of Control Parameters}

This subsection first  presents the second design strategy for the control parameters.  Then, it compares  both strategies from the perspectives of scalability and conservatism. 

\begin{remark}\label{rmk:designsteps}
Let Conditions \ref{ass:ABStabilizable}, \ref{ass:cnd3*}, and \ref{ass:spngtree} hold. Suppose that for any $i \in \mathcal{N}$, there is a solution to the LME \eqref{eq:newcondition}. 
Per Theorem \ref{thm:cnd3*tocnd3} and Corollary \ref{crl:maintheorem}, the first two steps in Remark \ref{rmk:designtrivial} remain unchanged, while the rest are updated as follows. 
\begin{enumerate}
    \item[(iii)] For each $i \in \mathcal{N}$, find a pair $(X_i,U_i)$ that satisfies the partition-independent regulator equations \eqref{eq:regulatoreqgeneral}.
    \item[(iv)] For each $i \in \mathcal{N}$, find a solution $Y_i$ to the LME \eqref{eq:newcondition}.
    \item[(v)] For each $i \in \mathcal{N}$, let $K_{2i} = \tilde{U}_i-K_{1i}\tilde{X}_i$ where $\tilde{X}_i$ and $\tilde{U}_i$ are as defined in \eqref{eq:sol_pair_tilde}.
\end{enumerate}
\end{remark}

\begin{remark}\label{rmk:design2scalability}
The second design strategy involves solving the partition-independent regulator equations \eqref{eq:regulatoreqgeneral} and the LME \eqref{eq:newcondition}. Neither of these equations includes the $k$-partition transformation term $s_i$. Therefore, once each follower finds solutions to the partition-independent regulator equations \eqref{eq:regulatoreqgeneral} and  the LME \eqref{eq:newcondition}, it can use these solutions to recompute the feedforward gain $K_{2i}$ whenever $s_i$ is changed. As a result, the design strategy in Remark \ref{rmk:designsteps} is much more scalable than the one in Remark \ref{rmk:designtrivial}, as illustrated in Example \ref{exp:numexscalability}.
\end{remark}

\begin{remark}\label{rmk:firstdesignbroader}
Theorem \ref{thm:cnd3*tocnd3} establishes that if the second design strategy is applicable, 
so is the first design strategy. 
Yet, the converse is not true. To see this,  consider the MAS described in Remark \ref{rmk:converse} and let $\mathcal{F} = 1$. Observe that the conditions in Remark \ref{rmk:designtrivial} are satisfied. However, we conclude from Remark \ref{rmk:converse} that 
neither step (iii) nor step (iv) in Remark \ref{rmk:designsteps} is feasible. Consequently, the first design strategy applies to a broader class of leader and follower dynamics. 
\end{remark}
 
Table \ref{tab:DesignStrategies} summarizes 
the differences between the first and second design strategies. %Under Conditions \ref{ass:ABStabilizable}--\ref{ass:spngtree}, 
We recommend adopting the second design strategy if Condition \ref{ass:cnd3*} holds and, for any $i\in \mathcal{N}$, the LME \eqref{eq:newcondition} has a solution; otherwise, adopting the first one.
%\textcolor{blue}{Table \ref{tab:DesignStrategies} summarizes the discussion on the design strategies.} 
%\vspace{-0.11 cm}
%\vspace{-0.1 cm}
\begin{table}[h] 
\centering 
\caption{Differences in Design Strategies}\label{tab:DesignStrategies}
% Overall table: 4 columns
\begin{tabular}{|M{1.6cm}|M{1.6cm} M{2.2cm}!{\vrule width 1.5pt}M{1.7cm}|}
\hline
\cellcolor{gray!8} \diagbox[width=2cm, innerleftsep=0.5pt, innerrightsep=0.5pt, height=0.8cm, font=\footnotesize\bfseries]{Differences}{Strategies} & \cellcolor{red!8} \textbf{First} & \cellcolor{blue!8} \textbf{Second} & \cellcolor{gray!8} \textbf{Conclusion} \\ \hline

\cellcolor{gray!8} \textbf{Conditions} 
& \multicolumn{2}{M{4.2cm}!{\vrule width 1.5pt}}{%
  \begin{minipage}[c][0cm][c]{\linewidth}
    \begin{tikzpicture}[overlay, remember picture]
      \fill[red!8] (-0.23,-0.58) rectangle (1.79cm,0.685cm);  % paint left half of the cell
    \end{tikzpicture}
    \noindent
    \hspace{-0.2cm}\makebox[0.4\linewidth][l]{Condition~\ref{ass:regeqsolution}}%
    \makebox[0.1\linewidth][c]{%
      \begin{tikzpicture}[overlay, remember picture]
      \fill[blue!8] (0.56,-0.58) rectangle (3.2cm,0.685cm);  % paint right half of the cell
    \end{tikzpicture}$\raisebox{-0.8ex}{$\underset{\text{(Remark \ref{rmk:converse})}}%
  {\overset{\text{(Theorem \ref{thm:cnd3*tocnd3})}}%
  {\overset{\scalebox{1}[1]{\ensuremath{\impliedby}}}{\rlap{\(\quad\not\)}\implies}}}$}$%
    }%
    \makebox[0.6\linewidth][r]{%
      \begin{tabular}{r}
        Condition \ref{ass:cnd3*} \vspace{-0.05cm} and \\ 
        solvability of \eqref{eq:newcondition} 
      \end{tabular}%
    }
  \end{minipage}
}
& \cellcolor{gray!8} The first is more general, as revealed in Remark \ref{rmk:firstdesignbroader}. \\ \hline

\cellcolor{gray!8} \textbf{LMEs to be solved}
& \cellcolor{red!8} Partition-Dependent:  \eqref{eq:regulatoreqtriv} 
& \cellcolor{blue!8} Partition-Independent:  $ \ \ \ \ $\eqref{eq:regulatoreqgeneral} and \eqref{eq:newcondition}
& \cellcolor{gray!8} The second is more scalable, as discussed in Remark \ref{rmk:design2scalability}. \\ \hline
\end{tabular}
\end{table} 
%\vspace{-0.3 cm}
%\vspace{-0.1 cm}

%\begin{remark}
%\textcolor{blue}{
%    The distributed observer in \eqref{eq:classicalcontroller} assumes all followers have access to the  matrix $A_0$. 
%To relax this assumption to a small subset of the followers, adaptive distributed observers (e.g., see \cite{cai2017adaptive})
%can be employed. In this case, the proposed second design strategy is even more crucial since the regulator equations to be solved become a matrix differential equation instead of the LME \eqref{eq:regulatoreqtriv}}.
%\end{remark}

\subsection{Solvability of the Introduced LME}

This subsection provides the straightforward characterizations of the solvability of the LME \eqref{eq:newcondition}.

\begin{proposition}\label{prp:solvabilityLMEs} Let $i \in \mathcal{N}$.
Then the following conditions are equivalent:
\begin{enumerate}
    \item[(i)] There exists a solution $Y_i$ to the LME \eqref{eq:newcondition}.  
    \item[(ii)]  The following inclusion holds:  
\begin{eqnarray}\label{eq:set_inclusion} \mathrm{im}
    \begin{bmatrix}
                E_i \\
                G_i 
            \end{bmatrix} \subseteq \mathrm{im}\begin{bmatrix}
                B_i \\
                D_i \\
            \end{bmatrix}.
            \end{eqnarray} 
\item[(iii)] The following rank condition holds: 
    \begin{eqnarray}\label{eq:necesssufLMEs}
            \mathrm{rank}\begin{bmatrix}
                B_i & E_i \\
                D_i & G_i 
            \end{bmatrix} = \mathrm{rank}\begin{bmatrix}
                B_i \\
                D_i
            \end{bmatrix}.
    \end{eqnarray}
\end{enumerate}
\end{proposition}

\begin{remark}
The inclusion \eqref{eq:set_inclusion} is a structural characterization of the solvability of the LME \eqref{eq:newcondition}. It is also easily testable by the rank condition \eqref{eq:necesssufLMEs}. 
The inclusion \eqref{eq:set_inclusion} holds if, and only if, follower $i$ is subject to solely matched disturbances.
%, which are common in certain applications \textcolor{blue}{(e.g., wheel slip in nonholonomic wheeled robots \cite{nonholowheelrobots1}, rotor imbalance in UAVs \cite{UAV1}, and aerodynamic drag in spacecrafts \cite{CANUTO2012121})}. 
Examples of such disturbances include input disturbances in wheeled robots \cite{nonholowheelrobots1}, %which can model wheel slip,
rotor imbalance in magnetic levitation systems \cite{10844942}, and the effect of wind on rotational motion of quadrotors \cite{Bayrak03042022}.
%\textcolor{blue}{Input disturbances in wheeled robots \cite{nonholowheelrobots1}, which can model wheel slip, rotor imbalance in magnetic levitation systems \cite{10844942}, and the effect of wind on rotational motion of a quadrotor \cite{Bayrak03042022}, are examples of matched disturbances.} 
Nevertheless, the first design strategy can still be employed  in the presence of unmatched disturbances. 
%\textcolor{blue}{In summary, apply the second design strategy if the LME (10) is solvable and apply the first design strategy otherwise.}
\end{remark}

\section{Experimental and Numerical Illustrations}\label{sec:numericalexample}

This section demonstrates the MORP's flexibility in shifting mission objectives via an experiment with networked mobile robots. It also provides two numerical examples to showcase the MORP's generality and compare the first and second design strategies regarding scalability. The following matrices are used throughout this section for the dynamics of the MASs.
\begin{comment}
Through \textcolor{blue}{an experiment and two numerical examples}, this section demonstrates the generality of the MORP and compares both design strategies regarding scalability. The following matrices are used to construct heterogeneous MASs in the examples
\end{comment}
\begin{eqnarray} \nonumber 
A_{\alpha} &=&  \begin{bmatrix}
0_{2 \times 2} & I_2\\
0_{2 \times 2} & 0_{2 \times 2} 
\end{bmatrix}, \ \Gamma  =  \begin{bmatrix}
0 & 0.0025  \\
-0.0025 & 0 
\end{bmatrix},  \nonumber \\ 
A_{\beta}  &=& \begin{bmatrix}
0.2 & 3 \\
0.1 & -0.1 \\
\end{bmatrix}, \ 
B_{\beta}  =  \begin{bmatrix}
0 & 3 \\
1 & 0
\end{bmatrix}. \nonumber 
\end{eqnarray}

\begin{comment}
\begin{eqnarray} \nonumber 
    A_{\alpha} &=& \begin{bmatrix}
0.2 & 3 \\
0.1 & -0.1 \\
\end{bmatrix}, \ 
B_{\alpha}  =  \begin{bmatrix}
0 & 3 \\
1 & 0
\end{bmatrix}, \\ \nonumber 
\textcolor{blue}{A_{\beta}} &\textcolor{blue}{=}&  \textcolor{blue}{\begin{bmatrix}
0_{2 \times 2} & I_2\\
0_{2 \times 2} & 0_{2 \times 2} 
\end{bmatrix}}, \
\textcolor{blue}{B_{\beta}  =  \begin{bmatrix}
0 \\
I_2
\end{bmatrix},} \ \textcolor{blue}{E = \begin{bmatrix}
0 \\
-4  
\end{bmatrix}}, \nonumber \\ \textcolor{blue}{D} &\textcolor{blue}{=}&  \textcolor{blue}{\begin{bmatrix}
0 & 0.01 
\end{bmatrix}, \ \Gamma  =  \begin{bmatrix}
0 & 0.0025  \\
-0.0025 & 0 
\end{bmatrix}.} \nonumber
\end{eqnarray}
\end{comment}

\begin{figure}[htbp]
  \begin{center}
\includegraphics[width=0.375\textwidth,trim=75 10 105 38,clip]{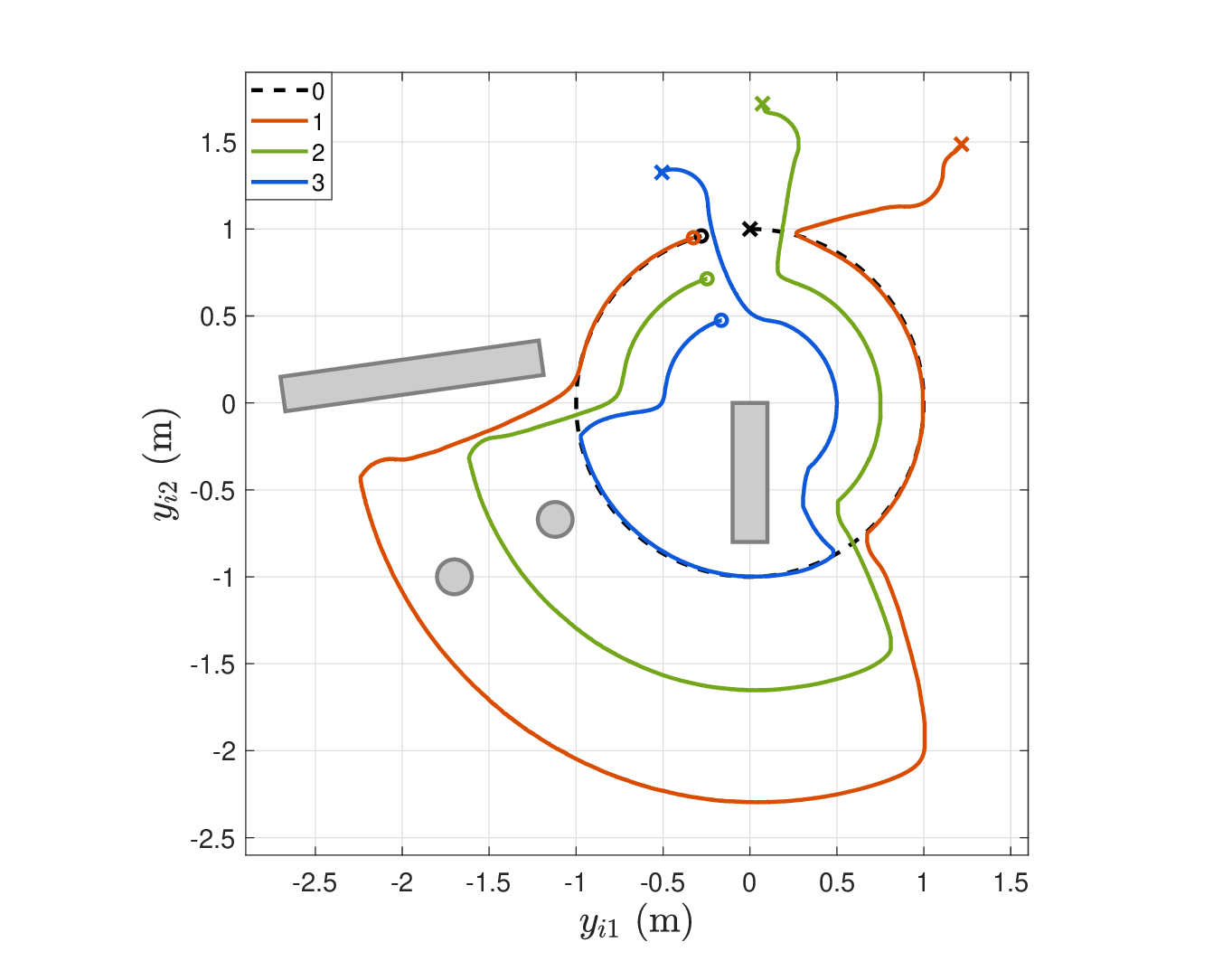} %\vspace{-0.1 cm}
    \caption{The trajectories of the MAS. Here, $y_{ij}$ denotes the $j$th entry of follower $i$'s output for $i=1,2,3$ and $y_{0j}$ denotes the $j$th entry of $v$ while ``$\times$" and ``o" marks $y_{ij}$ at the initial and final times for $i=0,1,2,3$, respectively.}
    %\vspace{-0.3 cm}
  \label{fig:numex_flexibility}
  \end{center}
\end{figure}

\begin{experiment}\label{exp:experiment}
%This example experimentally demonstrates that sequential objectives can be accomplished with the MORP.
In this experiment, a scenario with a MAS of $3$ nonholonomic mobile robots as followers operating in an adverse environment as first responders to an emergency is simulated in a laboratory setting. 
%In this experiment, a MAS of mobile robots tasked with a search and rescue mission in an adverse operating environment is simulated in a laboratory setting. 
The hand position dynamics of the followers (see Section II in \cite{1261347} for modeling details) and the leader are determined by the matrices: 
%To this end, consider a MAS composed of 3 nonholonomic mobile robots as followers with the hand position dynamics given by the matrices: 
$A_i = A_{\alpha}$, $B^\mathrm{T}_i = [0 \ I_2]$, $C_i = [I_2 \ 0 ]$, $D_i  = G_i = 0$, $E_i = 0$, $F_i = -I_2$ for $i = 1,2,3$;  $A_0 = \Gamma$. We take each mobile robot's hand position distance $0.15 \ \hspace{-0.5mm} \mathrm{m}$, mass $1 \ \hspace{-0.5mm} \mathrm{kg}$, and moment of inertia $0.01 \mathrm{\ \hspace{-0.5mm} kg \ \hspace{-0.5mm} m^2}$.
%(See \cite{1261347} for modeling details). 
The agents communicate over the augmented signed directed graph $\mathcal{G}(\Bar{\mathcal{A}})$, with $a_{21} = -1$, $a_{32} = 5$, and 
$f_1 = 1$ and the remaining entries of $\mathcal{A}$ and $\mathcal{F}$ are zero.

We assume that  cylindrical and cuboid  obstacles in the operating terrain are detected through a distributed sensor
fusion algorithm that runs onboard each robot. Accordingly, the robots update their $k$-partition transformation terms. To simulate this, the $k$-partition transformation is defined as a piecewise constant function. Specifically, $S(t) = \mathrm{diag}(1, 0.75, 0.5) $ for $t \in [0, 86.5) \cup [161, 212]$ seconds and $S(t) = \mathrm{diag}(2.3,1.65,1)$ for $t \in [86.5,161)$ seconds. Note that Conditions \ref{ass:ABStabilizable}, \ref{ass:cnd3*}, \ref{ass:spngtree}, and the inclusion \eqref{eq:set_inclusion} hold.
Following Remark \ref{rmk:designsteps}, we set $\mu = 10$ and design $K_{1i}$ using  $place$ function in MATLAB\footnote{For $i = 1,2,3$, $\mathrm{spec}(A_{i}+B_{i}K_{1i}) = \{ -0.75,-1.25,-1.75,-2.5\}$.}. As per steps (iii) and (iv), for $i= 1,2,3$, a solution pair $(X_i,U_i)$ to the partition-independent regulator equations \eqref{eq:regulatoreqgeneral} is recovered from their equivalent system of linear equations (see the proof of Theorem 1.9 in \cite{huang2004nonlinearbook}) and a solution $Y_i$ to the LME \eqref{eq:newcondition} is found using $linsolve$ function in MATLAB. %\footnote{The authors thank Jackson Kulik for his comment on solving \eqref{eq:newcondition}.}. 
Lastly, based on $S$,
%the $k$-partition transformation of interest,
we calculate $K_{2i}$ for each follower as in step (v).

%\textcolor{blue}{
%We assume that the MAS has the ability to identify no-drive zones in the operating terrain in a distributed manner and accordingly update the $k$-partition transformation $S$. Cylindrical and cuboid obstacles in the laboratory represent these zones. 
%The $k$-partition transformation is defined as a piecewise constant function to simulate the mentioned updates during the operation. Specifically, $S(t) = \mathrm{diag}(1, 0.75, 0.5) $ for $t \in [0, 86.5) \cup [161, 212]$ and $S(t) = \mathrm{diag}(2.3,1.65,1)$ for $t \in [86.5,161)$.
%Note that Conditions \ref{ass:ABStabilizable}, \ref{ass:cnd3*}, and \ref{ass:spngtree} hold. Following the second design strategy, we set $\mu = 10$ and design $K_{1i}$ using  \textcolor{blue}{$place$} function in MATLAB\textcolor{blue}{\footnote{\textcolor{blue}{For $i = 1,2,3$, $\mathrm{spec}(A_{i}+B_{i}K_{1i}) = \{ -0.75,-1.25,-1.75,-2.5\}$.}}}. As per steps (iii) and (iv), for $i= 1,2,3$, \textcolor{blue}{a solution pair $(X_i,U_i)$ to the partition-independent regulator equations \eqref{eq:regulatoreqgeneral} is recovered from their equivalent system of linear equations \textcolor{blue}{(see Theorem 1.9 in \cite{huang2004nonlinearbook})} and a solution $Y_i$ to the LME \eqref{eq:newcondition} is found using $linsolve$ function in MATLAB\footnote{\textcolor{blue}{We thank Dr. Jackson Kulik for his comments on solving \eqref{eq:newcondition}.}}.} Lastly, based on $k$-partition transformation of interest, we calculate $K_{2i}$ for each follower as suggested in step (v).}

The experiment is initiated with $x_1^{\mathrm{T}}(0) = [1.2, 1.5, 0, 0 ]$, $x_2^{\mathrm{T}}(0) = [0.1, 1.7, 0, 0]$, $x_3^{\mathrm{T}}(0) = [-0.5, 1.3, 0, 0]$, $\eta_i(0) = 0$ for $i = 1,2,3$, and $v^{\mathrm{T}}(0) = [0, 1]$. As seen in Fig. \ref{fig:numex_flexibility}, the output of follower $i$ tracks the $s_i$ multiple of the leader's state  
and steers around obstacles successfully for $i = 1, 2, 3$. The experiment video can be found at  \href{https://youtu.be/pUVTwtMqbP8}{https://youtu.be/pUVTwtMqbP8}.
%the experiment video can be found in the  \href{https://youtu.be/pUVTwtMqbP8}{clickable link}.
%each follower's output tracks the corresponding $k$-partition transformation term multiple of the leader's state
%as promised by Corollary \ref{crl:maintheorem}
%and steers around obstacles successfully. 
%The video record of the experiment can be found in the  \href{https://youtu.be/pUVTwtMqbP8}{clickable link}.
\end{experiment}

\begin{example}\label{exp:numexborp}
As indicated in Remark \ref{rmk:CompriseCORP}, the MORP includes the BORP. 
To make the MORP comparable with the existing solutions to the BORP, we force $k$-partition transformations to be gauge transformations. Despite such a restriction, this example presents the generality of the MORP.  
To this end,  consider $100$ followers and a leader with the following matrices: 
$A_i = A_\beta$, $B_i = B_{\beta}$, $C_i = [1 , 0]$ for $i = 1, \ldots, 50$; $A_i = A_{\alpha}$, 
$B^\mathrm{T}_i = [0 \ I_2]$, $C_i = [1, 0, 0, 0]$ for $i  =  51, \ldots, 100$; $D_i = 0$, $E_i = 0$, $G_i = 0$, $F_i = -1$ for $i = 1, \ldots, 100$; $A_0 = 0$. They communicate over 
%the augmented signed directed graph 
$\mathcal{G}(\Bar{\mathcal{A}})$ with $a_{i1} = 1$ for $i = 3,5, \ldots, 99$, $a_{i1} = -1$ for $i = 2, 4, \ldots, 100$, and $f_{1} = 1$, while the remaining entries of $\mathcal{A}$  and $\mathcal{F}$ are zero.

As the bipartite framework partitions the followers\footnote{There are studies incorporating the leader into the partition through
the structural balance condition on $\mathcal{G}(\Bar{\mathcal{A}})$ (e.g., see \cite{JiaoLewis}). This, however, allows only $1$ BORP to be solved.} based on the signs of the entries of $\mathcal{A}$, for the considered $\mathcal{G}({\mathcal{A}})$, it yields a unique $2$-partition of the followers. Therefore, the existing formulations in 
\cite{9089253, 8287187, huangbipartite} 
allow only $2$ BORPs to be solved by
swapping the followers that  track the leader's state and its additive inverse. On the other hand, as discussed in Section \ref{sec:arbitrarypartition},  the number of $2$-partitions of the followers and $1$-partition of the followers obtained by $k$-partition transformations are respectively $2^{99} - 1$ and $1$, which are the corresponding Stirling numbers of the second kind. 
In fact, there are $2^{100}$ gauge transformations generating all the aforementioned $2^{99}$ partitions. Thus, the proposed formulation allows $2^{100}$ BORPs to be solved 
without changing the underlying graph.

%The simulation is initiated with $v(0) = 1$, $\eta_i(0) = 0$ for $i = 1,\ldots, 100$, $x_i^{\mathrm{T}}(0) = [3i/50,  0]$ for $i = 1, \ldots, 50$, and $x_i^{\mathrm{T}}(0) = [-3i/50+ 3, 0, 0, 0]$ for $i =  51,\ldots, 100$.

Let $v(0) = 1$, $\eta_i(0) = 0$ for $i = 1,\ldots, 100$, $x_i^{\mathrm{T}}(0) = [3i/50,  0]$ for $i = 1, \ldots, 50$, and $x_i^{\mathrm{T}}(0) = [-3i/50+ 3, 0, 0, 0]$ for $i =  51,\ldots, 100$ in the simulations. 
%Let   $x_i^{\mathrm{T}}(0) = [3i/50,  0]$ for $i = 1, \ldots, 50$, $x_i^{\mathrm{T}}(0) = [-3i/50+ 3, 0, 0, 0]$ for $i =  51,\ldots, 100$, $\eta_i(0) = 0$ for $i = 1,\ldots, 100$, and $v(0) = 1$. 
The top row of Fig. \ref{fig:numexp_bipartite} illustrates the output responses of 2 BORPs that can be solved using the bipartite framework and the proposed formulation, where $S = \pm \left(I_{50} \otimes  \mathrm{diag}(1,-1)\right)$. 
For these BORPs, the design\footnote{There is a typo in Equation (4b) of \cite{9089253}. For Theorem 1 in \cite{9089253} to be valid, the term $(z_j-\mathrm{sgn}(a_{ij})z_i)$ needs to be replaced with $(\mathrm{sgn}(a_{ij})z_j-z_i)$.} in 
\cite{9089253}
and  the first and second design strategies can generate identical output responses. The bottom row of Fig. \ref{fig:numexp_bipartite} presents $2$ out of $2^{100} -2$ BORPs that can be solved with the proposed formulation but not with the  bipartite framework under the same graph. Here, $S = \pm \left( \mathrm{diag}(1,-1) \otimes I_{50}\right)$. 

%In the simulation, we take the initial states as $x_i^{\mathrm{T}}(0) = [3i/50,  0]$ for $i = 1, \ldots, 50$, \textcolor{blue}{$x_i^{\mathrm{T}}(0) = [-3i/50+ 3, 0, 0, 0]$} for $i =  51,\ldots, 100$, $\eta_i(0) = 0$ for $i = 1,\ldots, 100$, and $v(0) = 1$. The top row of Fig. \ref{fig:numexp_bipartite} illustrates the output responses of 2 BORPs that can be solved using the bipartite framework and the proposed formulation.
%For these BORPs, the design\footnote{There is a typo in Equation (4b) of \cite{9089253}. For Theorem 1 in \cite{9089253} to be valid, the term $(z_j-\mathrm{sgn}(a_{ij})z_i)$ needs to be replaced with $(\mathrm{sgn}(a_{ij})z_j-z_i)$.} in 
%\cite{9089253}
%and our design generate identical output responses. The bottom row of Fig. \ref{fig:numexp_bipartite} presents $2$ out of $2^{100} -2$ BORPs that can be solved with the proposed formulation but not with the existing bipartite framework.

\begin{figure}[h]
  \begin{center}
    \includegraphics[width=0.43\textwidth,trim=65 13 33 35,clip]{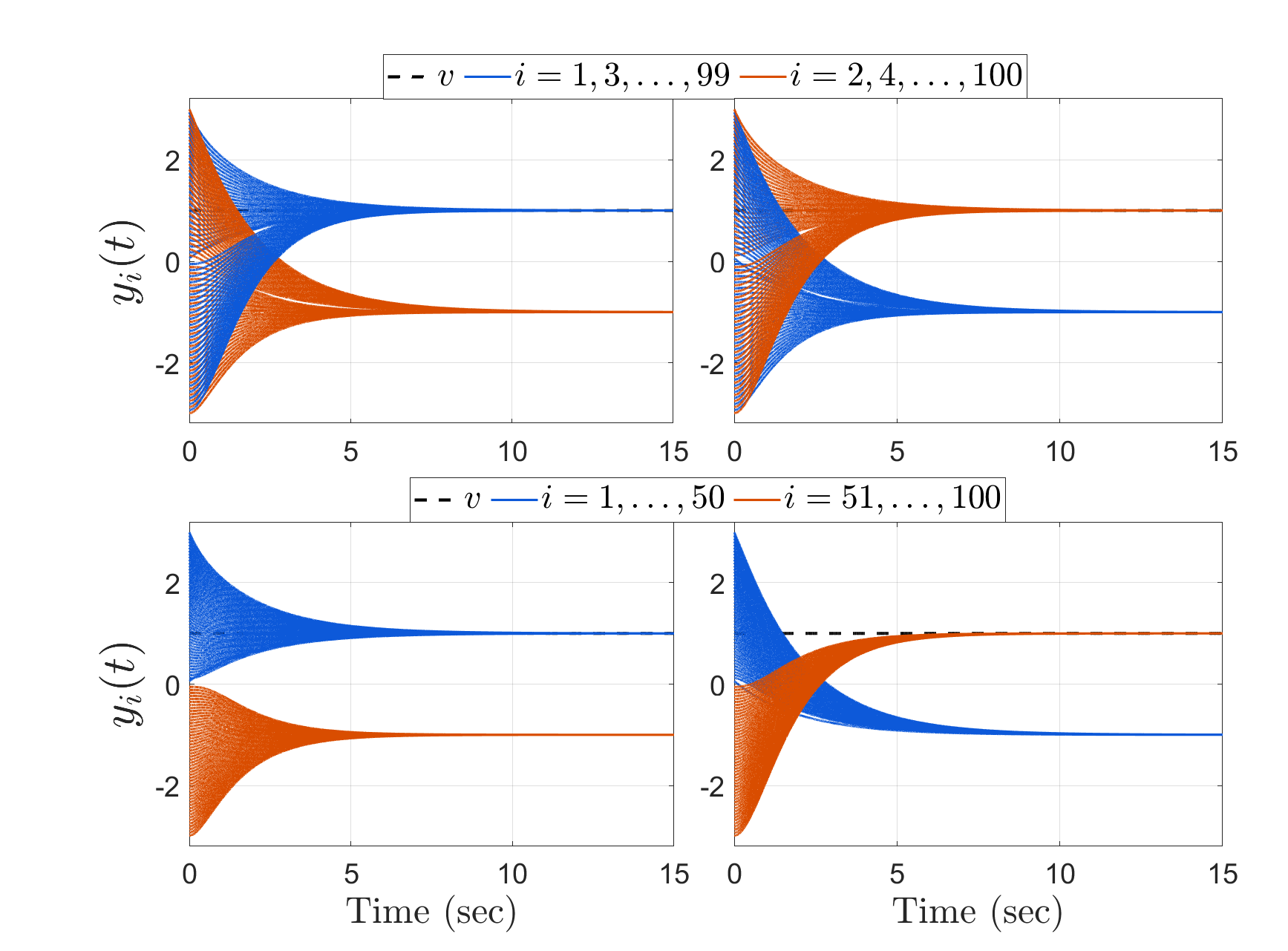} %\vspace{-0.1 cm}
    \caption{%The top row depicts the identical output responses with the design in \cite{9089253} and the proposed design. The bottom row presents the output responses with the proposed formulation for 2 BORPs that are impossible to formulate with the approach in \cite{9089253} without changing the underlying graph.
    %Top row: Output responses generated by both the design in \cite{9089253} and the proposed design, showing identical results. Bottom row: Output responses from the proposed formulation for 2 BORPs that cannot be formulated using the approach in \cite{9089253} without altering the underlying graph structure.
    The top row depicts the identical output responses with the design in 
    %\cite{JiaoLewis} 
    \cite{9089253}
    and the proposed design. The bottom row presents the output responses with the proposed formulation for 2 BORPs that are impossible to formulate with the approach in
    %with the formulation in 
    \cite{9089253}
    %\cite{JiaoLewis}
    without altering the underlying graph.} 
        %\vspace{-0.1 cm}
  \label{fig:numexp_bipartite}
  \end{center}
\end{figure}
\end{example}

\begin{example}\label{exp:numexscalability}
This example compares the partition-dependent steps of both design strategies in terms of scalability. In particular, Fig. \ref{fig:numexp_scalability} shows the total elapsed times with an average laptop for follower 1 in Example \ref{exp:numexborp} with $D_1 = [0 ,1]$ and $E^\mathrm{T}_1 = [0, -4]$ to complete steps (iii) and (iv) of the first design (see Remark \ref{rmk:designtrivial}) and  step (v) of the second design (see Remark \ref{rmk:designsteps}) 
as the cardinality of the given set\footnote{Such sets are generated using  $randn$ function in MATLAB.} of $k$-partition transformation terms increases.
With the first design strategy, feedforward gains for up to $190$ $k$-partition transformation terms can be computed within $4$ milliseconds.
On the other hand, with the second one, feedforward gains for approximately $10000$ $k$-partition transformation terms can be computed within the same amount of time. 

\begin{figure}[t]
  \begin{center}
    \includegraphics[width=0.4\textwidth,trim=40 6 70 0,clip]{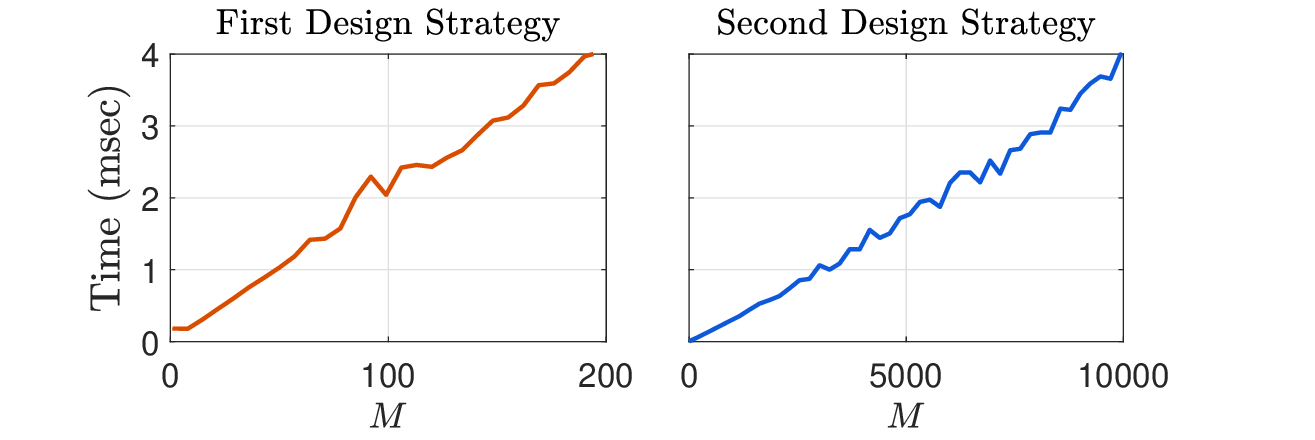} 
   % \vspace{-0.1 cm}
    \caption{Elapsed times of both design strategies with respect to the cardinality $M$ of the given set of $k$-partition transformation terms.
    }  
    %\vspace{-0.31 cm}
  \label{fig:numexp_scalability}
  \end{center}
\end{figure}
\end{example}

\section{Conclusion}
The primary motivation of this article has been to provide MASs with objectives beyond cooperation and bipartition for tactical flexibility in adverse operating environments. To this end, the MORP for linear MASs has been formulated and solved for the first time. Two design strategies for the control parameters have been proposed. The first applies to a broader set of MASs, but it has a drawback  due to the partition-dependent regulator equations. 
%suffering from a drawback  (i.e., scalability).
The second eliminates this drawback, and hence, it is significantly more scalable, yet applicable if the followers are subject to solely matched disturbances. Table \ref{tab:DesignStrategies} summarizes
the differences between the first and second design strategies. 
Theoretical results have been demonstrated by 
experimental and numerical tests.
%Experimental and numerical studies confirm the theoretical results.
%the discussion on the proposed design strategies' scalability and conservatism. 
%The distributed observer in \eqref{eq:classicalcontroller} assumes all followers have access to the  matrix $A_0$. 
%To relax this assumption to a small subset of the followers, 
%solving the MORP with distributed control laws involving adaptive distributed observers (e.g., see \cite{cai2017adaptive})
%would be well worth an exploration. \textcolor{blue}{In this case, the proposed second design strategy becomes even more crucial since follower $i$ has to recompute a solution to a matrix differential equation instead of \eqref{eq:regulatoreqtriv} whenever the $s_i$ is updated}. 
A future research direction to widen the application domain of the MORP is $k$-partition transformation generation as a distributed optimization problem, where the followers optimize a MAS-level objective such as total fuel or energy consumption. For robustness, studying the MORP with an internal model-based distributed control law is also worth exploring. Lastly, merging the MORP with the cluster problem is another research topic, where each cluster solves its own MORP. %\textcolor{blue}{Solving the robust MORP with an internal model-based distributed control law is another research direction.} 

\bibliographystyle{IEEEtran}
\bibliography{refs.bib}

\section*{Appendix}

\begin{lemma}\label{lmm:muvalue}
Suppose Condition \ref{ass:spngtree} holds. 
          The matrix $A_{\mu}$ is Hurwitz if, and only if, $\mu$ satisfies the inequality \eqref{eq:muselection}. 

\end{lemma}
\begin{IEEEproof}
    All the eigenvalues of $A_{\mu}$ are as follows: 
\begin{eqnarray}
    \lambda_j(A_0) - \mu \lambda_i(\mathcal{H}), \quad j = 1, \ldots, n_0, \  i = 1, \ldots, N  \nonumber 
\end{eqnarray}
(see the proof of Theorem 1 in \cite{su2012cooperative}). 
One can use this fact and 
Lemma 1 in \cite{su2012cooperative} to conclude the proof.  
\end{IEEEproof}

\end{document}